\documentclass[twocolumn,showpacs,preprintnumbers,aps,pra,amsmath,amssymb]{revtex4}

\usepackage{graphicx}
\usepackage{dcolumn}
\usepackage{bm}
\usepackage{amssymb}
\usepackage[german, english]{babel}
\begin{document}
\draft

\title{Lifetime of angular momentum in a rotating strongly interacting Fermi gas}

\author{S. Riedl,$^{1,2}$ E. R. S\'{a}nchez
Guajardo,$^{1,2}$ C. Kohstall,$^{1,2}$ J. Hecker Denschlag,$^{1}$
and R. Grimm$^{1,2}$}

\address{$^{1}$Institut f\"ur Experimentalphysik und Zentrum f\"ur Quantenphysik,
Universit\"at Innsbruck, 6020 Innsbruck, Austria\\
$^{2}$Institut f\"ur Quantenoptik und Quanteninformation,
\"Osterreichische Akademie der Wissenschaften, 6020 Innsbruck,
Austria}

\date{\today}

\pacs{67.25.dg, 05.30.Fk, 67.85.Lm, 34.50.Cx}

\begin{abstract}
We investigate the lifetime of angular momentum in an ultracold
strongly interacting Fermi gas, confined in a trap with controllable
ellipticity. To determine the angular momentum we measure the
precession of the radial quadrupole mode. We find that in the
vicinity of a Feshbach resonance the deeply hydrodynamic behavior in
the normal phase leads to a very long lifetime of the angular
momentum. Furthermore, we examine the dependence of the decay rate
of the angular momentum on the ellipticity of the trapping potential
and the interaction strength. The results are in general agreement
with the theoretically expected behavior for a Boltzmann gas.

\end{abstract}

\maketitle

\section{Introduction}

The dynamics of an ultracold quantum gas is an important source of
information on the physical nature of the system. A particularly
interesting situation is an atomic Fermi gas in the vicinity of a
Feshbach resonance \cite{Inguscio2006ufg,Giorgini2008tou}. The
Feshbach resonance allows us to tune the two-body interaction and
thus to control the coupling between the atoms. It connects a
molecular Bose-Einstein condensate (BEC) with a
Bardeen-Cooper-Schrieffer (BCS) superfluid. In the crossover region
between these two limiting cases the center of the Feshbach
resonance is of special interest. Here the unitarity-limited
interactions lead to universal behavior of the Fermi gas.

The strong two-body interactions close to the Feshbach resonance
lead to very low viscosity and hydrodynamic behavior in the normal
phase, similar to properties of a superfluid
\cite{Clancy2007oon,Wright2007ftc}. The coexistence of normal and
superfluid hydrodynamic behavior is a special property of the
strongly interacting Fermi gas, which stands in contrast to
ultracold Bose gases, where deep hydrodynamic behavior is usually
restricted to the superfluid condensate fraction. The low-viscosity
hydrodynamic behavior leads to a long lifetime of collective motion
in the system. Using collective modes the dynamics has been
investigated in a broad range of temperatures and interaction
strengths in the crossover region
\cite{Clancy2007oon,Wright2007ftc,Bartenstein2004ceo,Kinast2004efs,Kinast2004boh,Kinast2005doa,Altmeyer2007pmo,Altmeyer2007doa,Riedl2008coo},
including the hydrodynamic regime in the normal phase. Another
important collective motion is the rotation of the gas, which is of
particular interest in relation to superfluidity
\cite{Zwierlein2005vas}.

In this Article, we study the lifetime of the angular momentum of a
rotating strongly interacting Fermi gas. We determine the angular
momentum using the precession of the radial quadrupole mode. This
method is well established to study the angular momentum in
experiments with BEC
\cite{Chevy2000mot,Haljan2001uos,Leanhardt2002ivi}. We observe that
the unique hydrodynamic behavior of the strongly interacting Fermi
gas leads to particularly long lifetimes of the angular momentum. We
perform a quantitative analysis of the dissipation of the angular
momentum caused by the trap anisotropy for a gas in the unitarity
limit. The measurements show general agreement with the expected
behavior for a Boltzmann gas \cite{Odelin2000sua}. As shown in a
previous study comparing experiment and theory \cite{Riedl2008coo},
a Boltzmann gas describes the behavior of a gas in the normal state
with unitarity-limited interactions reasonably well. Finally we
study the dependence of the lifetime on the interaction strength of
the gas in the crossover region between the BEC and BCS regime.



\section{Experimental procedure}

To realize an ultracold strongly interacting Fermi gas we trap and
cool an equal mixture of $^6$Li atoms in the lowest two atomic
states as described in our previous work
\cite{Jochim2003bec,Altmeyer2007doa}. We control the interparticle
interaction by changing the external magnetic field in the vicinity
of a broad Feshbach resonance centered at $834$\,G
\cite{Bartenstein2005pdo}. The atoms are held by an optical dipole
trap using a red-detuned, single focused laser beam and an
additional magnetic trap along the beam; this magnetic confinement
dominates over the optical confinement along the beam under the
conditions of the present experiments. The resulting trap provides
weak confinement along the beam ($z$ axis) and stronger transverse
confinement ($x$-$y$ plane), leading to a cigar-shaped cloud. The
trap is well approximated by a harmonic potential with trap
frequencies $\omega_x\approx\omega_y\approx2\pi\times800$\,Hz and
$\omega_z=2\pi\times25$\,Hz. The trap in general also has a small
transverse ellipticity, which can be controlled during the
experiments. We define an average transverse trap frequency as
$\omega_r=\sqrt{\omega_x\omega_y}$. The Fermi energy of the
noninteracting gas is given by
$E_F=\hbar(3N\omega_x\omega_y\omega_z)^{1/3}=\hbar^2k_F^2/2M$ where
$N=5\times10^5$ is the total atom number, $M$ is the atomic mass and
$k_F$ is the Fermi wave number. The corresponding Fermi temperature
is $T_F=E_F/k=1.3\,\mu$K, with $k$ the Boltzmann constant. The
interaction strength is characterized by the dimensionless parameter
$1/k_Fa$, where $a$ is the atomic $s$-wave scattering length.

To dynamically control the shape of the trapping potential in the
transverse plane we use a rapid spatial modulation of the trapping
laser beam by two acousto-optical deflectors, which allows us to
create time-averaged trapping potentials \cite{Altmeyer2007doa}. The
control over the potential shape has two different applications for
the measurements. As a first application we use it to adjust the
static ellipticity $\epsilon =
(\omega_x^2-\omega_y^2)/(\omega_x^2+\omega_y^2)$ of the trap in the
$x$-$y$ plane. This allows us to compensate for residual ellipticity
of the trapping potential, i.e.\ of the trapping laser beam, and
also to induce a well defined ellipticity. The second application is
the creation of a rotating elliptic potential with a constant
ellipticity $\epsilon'$ \cite{rotatingframe}. This is needed to spin
up the gas. Both the static ellipticity in the $x$-$y$ plane and the
rotating elliptic potential can be controlled independently. To
determine the ellipticity we measure the frequency of the sloshing
mode along the two principal axes of the elliptic potential. This
allows controlling the ellipticity with an accuracy down to
typically $0.005$.


\begin{figure}
  \includegraphics[width=8cm]{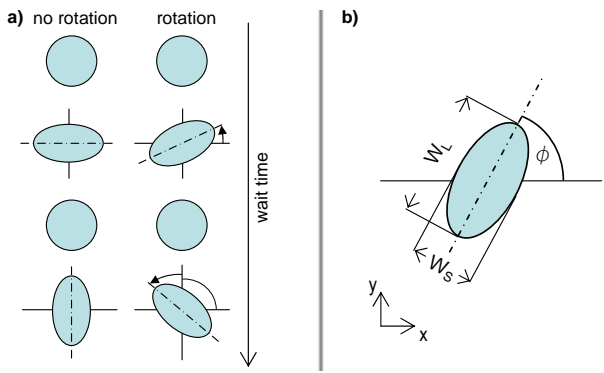}\\
  \caption{Oscillation of the cloud after excitation of the radial
  quadrupole mode. For a rotating hydrodynamic gas the principal axes of the
  quadrupole mode oscillation precess with a frequency determined by
  the angular momentum of the gas. To follow the precession we
  measure the angle of the long axis of the cloud. Note that every
  half oscillation period this angle changes by $\pi/2$ because of
  the mode oscillation; see also Fig.\,\ref{oscillation}. The
  oscillation of the cloud shape is determined by measuring the
  widths along the short ($W_S$) and the long axis ($W_L$) of the
  cloud. }\label{schematic}
\end{figure}


\begin{figure}
  \includegraphics[width=8cm]{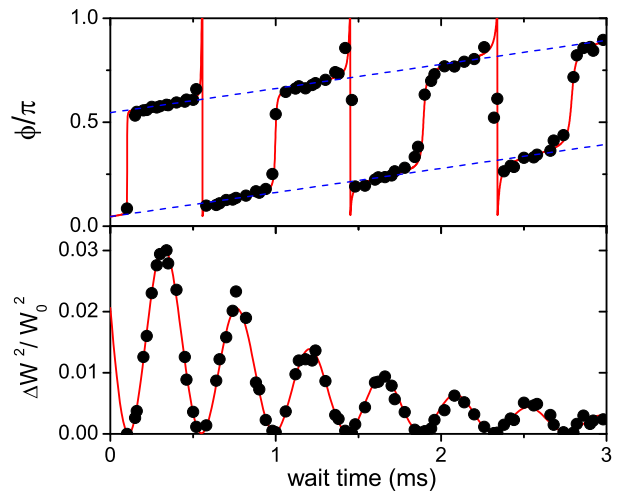}\\
  \caption{Evolution of the quadrupole mode in a rotating Fermi gas
  in the unitarity limit. The upper panel shows the precession of
  the principal axes of the mode. The experimental data are shown by
  the dots. The solid line represents a fit according to
  Eq.~\ref{fitprecession}. The dashed lines correspond to the
  idealized precession of the angle when there is no damping present
  in the mode. Whenever the oscillation of the difference in widths
  $\Delta W^2/W_0^2$ (lower panel) has a local maximum the observed
  precession angle coincides with the idealized precession. The
  parameter $W_0$ is the average width of the cloud. The finite
  value of $\phi$ at zero wait time results from the precession of
  the cloud during expansion. Here $L_z=1.7\hbar$ and
  $T/T_F\approx0.2$. }\label{oscillation}
\end{figure}

To measure the angular momentum of the cloud we exploit the fact
that collective excitation modes are sensitive to the rotation of
the cloud. Here we use the precession of the radial quadrupole mode
to determine the angular momentum of the rotating cloud; see
Fig.~\ref{schematic}. This method works under the general condition
that the gas behaves hydrodynamically \cite{Chevy2003kmo}. In our
case of a strongly interacting Fermi gas, this method probes both
the superfluid and the classically hydrodynamic part and does not
distinguish between these two components. For the case of atomic
BEC, the precession has been well studied in theory
\cite{Sinha1997sao,Dodd1997eso,Svindzinsky1998nmo,Zambelli1998qva}
and used in experiments to determine the angular momentum of the BEC
\cite{Chevy2000mot,Haljan2001uos,Leanhardt2002ivi}. For an atomic
BEC the non-condensed part is usually collisionless and does not
contribute to the mode precession.


The radial quadrupole mode consists of two collective excitations
with angular quantum numbers $m = +2$ and $m = -2$ and frequencies
$\omega_+$ and $\omega_-$, respectively. These two excitations
correspond to an elliptic deformation of the cloud rotating in
opposite directions. The superposition of the excitations results in
the radial quadrupole mode. For a gas at rest the two excitations
are degenerate, while for a gas carrying angular momentum the
frequencies are different, which causes a precession of the mode,
see Fig.~\ref{schematic}. The mode precesses with a frequency
$\Omega_\phi=(\omega_+-\omega_-)/4$. The angular momentum itself can
be calculated from the precession frequency \cite{Zambelli1998qva}
using
\begin{equation}\label{maineq}
    \Omega_\phi= L_z/(2 M r_{\rm rms}^2).
\end{equation}
Here $L_z$ is the average angular momentum per atom and $r_{\rm
rms}^2$ is the mean value of $x^2+y^2$ of the density distribution
\cite{rmswidth}.

To excite the quadrupole mode we switch on an elliptic potential for
$50\,\mu$s; this short elliptic deformation does not affect the
angular momentum of the gas. For the excitation we make sure that
$\omega_r$ does not change. This ensures that no compression mode is
excited and only an equal superposition of the $m=\pm2$ modes is
created \cite{Altmeyer2007doa}.

To follow the quadrupole oscillation we determine the angle of the
long axis, $\phi$, and the difference of the widths along the
principle axes of the cloud, $\Delta W=W_L-W_S$, after a variable
wait time in the trap; see Fig.~\ref{schematic}. Therefore we fit a
zero temperature, two-dimensional Thomas-Fermi profile to absorption
images \cite{fitdensitydistr}. We also keep the angle of the long
axis a free fit parameter. The width of the cloud is defined as
twice the Thomas-Fermi radius.

To resolve the density distribution in the $x$-$y$ plane we let the
cloud expand for $0.8$\,ms before taking the image. The expansion
does not only increase the width of the cloud but also leads to an
increase of the precession angle as a consequence of the angular
momentum. A quantitative analysis of the small contribution to the
total precession angle that results from the expansion is given in
Appendix B.

Figure \ref{oscillation} shows the evolution of the precessing
quadrupole mode. The upper part shows the precession angle. The
finite value of $\phi$ at zero wait time results from the expansion.
The periodic jumps of the precession angle reflect the alternation
between the long and the short axis while the quadrupole mode
evolves. As the precession proceeds, these jumps become more and
more smooth. This is caused by stronger damping of the $m=-2$
excitation compared to the $m=+2$ excitation. Similar behavior has
been observed in Ref.\,\cite{Bretin2003qoo} for the case of a BEC.
There the authors discuss two possible mechanisms where the
difference in damping is either due to a rotating thermal cloud
\cite{Williams2002dio} or Kelvin mode excitations
\cite{Chevy2003kmo}. From our measurements we cannot discriminate
between these two mechanisms.

To fit the observed precession of the quadrupole mode we use the
function given in Appendix A. We find very good agreement between
the data and the expected behavior. For the data set shown in
Fig.~\ref{oscillation} the angular momentum is $1.7 \hbar$. The
average damping rate is
$(\Gamma_-+\Gamma_+)/2=(460\pm30)$\,s$^{-1}$, while the difference
in the damping rate of the $m=-2$ compared to the $m=+2$ excitation
is $\Gamma_--\Gamma_+=(80\pm40)$\,s$^{-1}$.


We find that a simplified procedure can be used to determine the
angular momentum from a single measurement, instead of fitting the
whole precession curve. If the measurement is taken at a time when
$\Delta W^2$ has a local maximum, the precession angle $\phi$ is
independent of the distortion caused by the difference in the
damping rates between the two excitations; see
Fig.~\ref{oscillation}. This allows us to determine the difference
$\omega_+-\omega_-=4\,\phi/\Delta t $ and therefore to determine
$L_z$ with a single measurement. The duration $\Delta t$ is the sum
of the wait time in the trap and an effective precession time $t_e$,
which accounts for the precession of the quadrupole mode during
expansion as discussed in Appendix B. Depending on the damping of
the mode oscillation we measure the precession angle at the first or
second maximum of $\Delta W^2$ \cite{modefreqrotation}.

To determine the temperature of the gas in the unitarity limit we
first adiabatically change the magnetic field to $1132$\,G
\cite{technical}, where $1/k_Fa \approx -1$, to reduce the effect of
interactions on the density distribution \cite{Luo2007mot}. Under
this condition, for $T>0.2T_F$, the interaction effect on the
density distribution is sufficiently weak to treat the gas as a
noninteracting one and to determine the temperature from
time-of-flight images. We fit the density distribution after 2\,ms
release from the trap to a finite-temperature Thomas--Fermi profile.
The temperature measured at $1132$\,G is converted to the
temperature in the unitarity limit under the assumption that the
conversion takes place isentropically, following the approach of
Ref.~\cite{Chen2005toi}.

\section{Spinning up the gas}\label{secspin}
\begin{figure}
  \includegraphics[width=8cm]{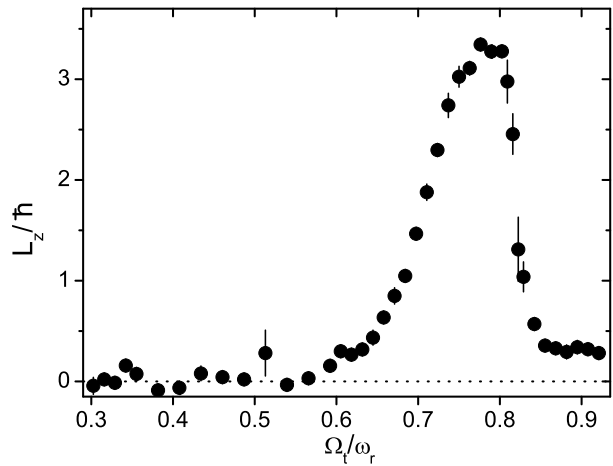}\\
  \caption{The angular momentum $L_z$ as a function of the rotation
  frequency $\Omega_t$ of the elliptic trap. Here we spin up the gas
  for $t_{\rm rot}=60$\,ms.  The temperature is $T/T_F\approx0.2$.
  The gas is in the unitarity limit. }\label{spinup}
\end{figure}
To spin up the gas we introduce a rotating anisotropy into the
initially round trap in the $x$-$y$ plane. More specifically, we
suddenly switch to a rotating elliptic trap potential with a
rotation frequency $\Omega_t$ and ellipticity $\epsilon'=0.03$,
rotate for a time $t_{\rm rot}$ on the order of $100$\,ms, and then
ramp down the ellipticity in $50$\,ms while the trap is still
rotating.

In the case of hydrodynamic behavior of the gas this spinning up
method is resonantly enhanced in a certain range of rotation
frequencies; see Fig.~\ref{spinup}. The reason for this behavior is
the resonant excitation of quadrupolar flow which leads to a dynamic
instability when $\Omega_t$ is close to half the oscillation
frequency of the radial quadrupole mode $\omega_q/2 =
0.71\,\omega_r$. This effect was used to nucleate vortices in a BEC
\cite{Madison2000vfi} and was further studied in
Refs.\,\cite{Madison2001sso,Hodby2002vni}. A signature of the
resonant excitation is a strong elliptic deformation of the cloud
shape which exceeds the ellipticity of the trap $\epsilon'$ during
the spin-up process. We clearly see this effect when we spin up the
gas. We also find that the rotation frequency where $L_z$ starts to
increase strongly depends on $\epsilon'$ and $t_{\rm rot}$ in a
similar way as it was observed in
Refs.\,\cite{Madison2001sso,Hodby2002vni}. Note that we cannot draw
any conclusion concerning superfluidity from the resonant behavior
of $L_z$ in Fig.~\ref{spinup} because it is only a consequence of
hydrodynamic behavior and the strongly interacting gas is
hydrodynamic both below and above $T_c$. In fact, for temperatures
clearly above $T_c$ we find similar behavior for $L_z$ as a function
of $\Omega_t$.


For an atomic BEC, $L_z$ was found to first increase abruptly from
$0$ to $1\hbar$ with $\Omega_t$, caused by the appearance of a
centered vortex \cite{Chevy2000mot}. As the formation of pairs is
necessary for superfluidity in the BEC-BCS crossover regime, the
angular momentum per atom of a single vortex in the center of the
cloud amounts to $L_z=\hbar/2$. We do not observe such an abrupt
increase of $L_z$. Nevertheless this does not exclude that vortices
are created during our spin-up process; the abrupt change of $L_z$
is not a necessary consequence of the creation of vortices as the
angular momentum of a vortex depends on its position in an
inhomogeneous gas \cite{Chevy2000mot}. Furthermore our measurement
of $L_z$ cannot distinguish between the angular momentum carried by
the superfluid and the normal part of the cloud. Also we cannot
directly observe vortices in our absorption images; we believe that
the reason is the very elongated cloud which strongly decreases the
contrast of the vortex core in the absorption images.

During our spin-up process we observe a significant heating of the
gas depending on the rotation frequency and the rotation time. We
keep these two parameters as small as possible. We find that a
rotation frequency of $\Omega_t/\omega_r=0.6$ and $t_{\rm
rot}=200$\,ms lead to an angular momentum of about $L_z=2\hbar$.
This is sufficient to perform the measurements, and at the same time
does only moderately increase the temperature.

We determine the temperature of the gas after the spin-up process.
To avoid complications in the temperature measurement we wait until
the rotation has completely decayed. To keep this wait time short,
on the order of $100$\,ms, we speed up the decay by increasing the
ellipticity of the trap; see discussion below. Note that the low
initial angular momentum used in the experiments, always staying
below $3\hbar$, does not lead to a significant increase in the
temperature when the rotation energy is completely converted into
heat \cite{tempincrease}.

\section{Lifetime of the angular momentum}
In an elliptic trap the angular momentum is not a conserved quantity
and hence can decay. The dissipation of $L_z$ is due to friction of
the gas caused by the trap anisotropy. Here we investigate the
dependence of the decay of $L_z$ on the static ellipticity for the
case of unitarity-limited interactions. We compare our experimental
results to the predicted behavior for a rotating Boltzmann gas
\cite{Odelin2000sua}. Finally we study the dependence of the decay
rate on the interaction strength in the BEC-BCS crossover regime.

The fact that the gas consists of two different components, the
normal and the superfluid part, leads in general to a complex
behavior for the decay of $L_z$. For example, in the case of a BEC
an exponential decay is related to the co-rotation of the thermal
cloud with the condensate \cite{Zhuravlev2001ddo,Abo-Shaeer2002fad}.
When the thermal cloud is not rotating, theoretical
\cite{Zhuravlev2001ddo} and experimental \cite{Madison2000vfi}
studies show nonexponetial behavior. For a gas completely in the
hydrodynamic regime it is expected that the decrease in $L_z$ has an
exponential form \cite{Odelin2000sua}.

To measure the decay rate of the angular momentum we use the
following procedure. After spinning up the gas as discussed in
Sec.~\ref{secspin}, we slowly increase the static ellipticity within
$10$\,ms, wait for a certain hold time to let the angular momentum
partially decay and then we remove the ellipticity again within
$10$\,ms. Finally we excite the radial quadrupole mode and observe
the precession to determine $L_z$ using the simplified procedure
discussed earlier.
\begin{figure}
    \includegraphics[width=8cm]{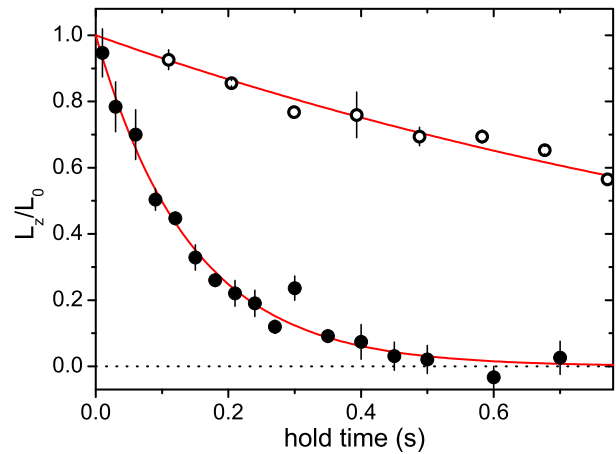}\\
  \caption{Decay of the angular momentum $L_z$ for a gas in the
  unitarity limit. The temperature is $T/T_F=0.22(3)$. We fit an
  exponential decay behavior (solid lines) to the experimental data
  points. For low ellipticity $\epsilon=0.009$ (open dots) the
  lifetime is $1.4$\,s, while at higher ellipticity $\epsilon=0.1$
  (filled dots) the lifetime is only $0.14$\,s. To better see the
  difference of the lifetime for the two ellipticities we normalized
  $L_z$ by its initial value $L_0$. For the lower ellipticity
  $L_0=2.2\hbar$ and for the higher ellipticity $1.6\hbar$.}
  \label{expdecay}
\end{figure}

In Figure \ref{expdecay} we show two examples for the decay of
$L_z$. We find that the decay of the angular momentum perfectly fits
an exponential behavior for all the static ellipticities,
temperatures, and interaction strengths we used. For the lowest
temperatures obtained the lifetime for a gas in the unitarity limit
goes up to $1.4$\,s, presumably limited by a residual anisotropy of
the trap. This lifetime is by more than a factor of thousand larger
then the radial trap oscillation period. Furthermore the lifetime of
the angular momentum is much larger than the lifetime of collective
excitation modes. For example the lifetime of the radial quadrupole
mode under the same conditions is only $2$\,ms. A larger ellipticity
of the trap significantly decreases the lifetime of $L_z$.

\begin{figure}
  \includegraphics[width=8cm]{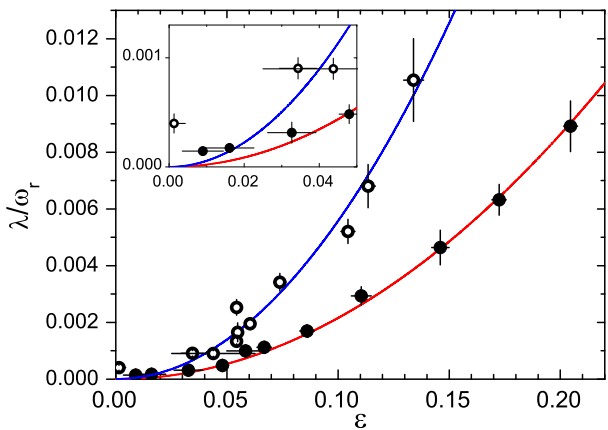}\\
  \caption{Normalized decay rate of the angular momentum as a
  function of the ellipticity for a gas in the unitarity limit. The
  temperatures are $T/T_F=0.22(3)$ (filled dots) and $0.35(2)$ (open
  dots). The solid lines are fits based on the expected behavior for a
  Boltzmann gas \cite{Odelin2000sua}. The inset shows the low
  ellipticity region. }\label{lifetimeellipticity}
\end{figure}

In the following we investigate quantitatively the dependence of the
decay rate of the angular momentum, $\lambda$, on ellipticity and
temperature. The experimental results are shown in
Fig.~\ref{lifetimeellipticity} for two different temperatures. The
full circles display the data for a temperature of $T/T_F=0.22(3)$
and the open circles correspond to a temperature of $T/T_F=0.35(2)$.
For better comparison with theory we plot the normalized decay rate
$\lambda/\omega_r$. A strong increase of the decay rate with
increasing ellipticity shows the important role of the trap
anisotropy on the lifetime of the angular momentum. For both
temperatures the qualitative behavior of the decay rate is the same.

Next we compare the behavior of the decay rate with a theoretical
prediction for a Boltzmann gas \cite{Odelin2000sua}. As we showed
recently in Ref.~\cite{Riedl2008coo}, a Boltzmann gas describes the
behavior of a unitarity-limited gas in the normal state reasonably
well. The predicted behavior of the decay rate is given by
$\lambda/\omega_r = 2\epsilon^2\omega_r\tau$ under the assumption
that $\epsilon \ll 1/(4 \omega_r\tau)$ \cite{condition}, where
$\tau$ is the relaxation time or effective collision time
\cite{Riedl2008coo,Vichi2000cdo,Huang1987xxx}. This condition is
well fulfilled in our system because the gas is in the hydrodynamic
regime where $\omega_r \tau\ll 1$. We compare this theoretical
prediction, with $\tau$ as a free parameter, to our measurements. We
find $\omega_r\tau=0.108(5)$ for the lower temperature and
$\omega_r\tau=0.28(1)$ for the higher temperature data.

Note that at very low ellipticity, $\epsilon<0.02$´, the observed
decay rate for both temperatures lies significantly above the
expected behavior; see inset of Fig.~\ref{lifetimeellipticity}. We
attribute this to an additional anisotropy of the trap beyond simple
ellipticity. This weak anisotropy becomes relevant only at very low
$\epsilon$. Furthermore the finite linear heating rate of the
trapped gas of $0.05\,T_F$\,s$^{-1}$ becomes important when the
decay rate is very low, which means that the lifetime of $L_z$ is on
the order of seconds. In this case the temperature cannot be assumed
to be constant during the decay of $L_z$.

A recent calculation of the relaxation time $\tau$ for a Fermi gas
in the unitarity limit \cite{Riedl2008coo} allows us to compare the
experimental values for $\omega_r\tau$ to theory. For $T/T_F=0.35$
the obtained relaxation time of $\omega_r\tau=0.28$ is clearly
larger than the calculated value of $\omega_r\tau=0.13$. This means
that the theory predicts that the gas is somewhat deeper in the
hydrodynamic regime compared to the experimental findings. Similar
deviations showed up when the theory was compared to the temperature
dependence of collective oscillations \cite{Riedl2008coo}. For the
lower temperature the obtained value for $\omega_r\tau$ cannot be
compared to the calculation of Ref.~\cite{Riedl2008coo} as the
theory is restricted to higher temperatures.

Finally we study the decay of the angular momentum in the crossover
region between the BEC and BCS regimes. We measure the decay rate
for different interaction parameters $1/k_Fa$. The experimental
sequence is the same as for the decay rate in the unitarity limit
beside ramping the magnetic field to the desired value in $100$\,ms
before increasing the ellipticity and ramping back the magnetic
field in $100$\,ms before exciting the quadrupole mode. Here the
magnetic field is changed slowly such that the gas is not
collectively excited. The ellipticity for all magnetic fields is set
to be $\epsilon=0.09$. This sizeable value of $\epsilon$ ensures
that a small anisotropy beyond ellipticity does not affect the decay
rate and makes the measurement less sensitive to heating while the
angular momentum damps out as discussed above.

\begin{figure}
  \includegraphics[width=8cm]{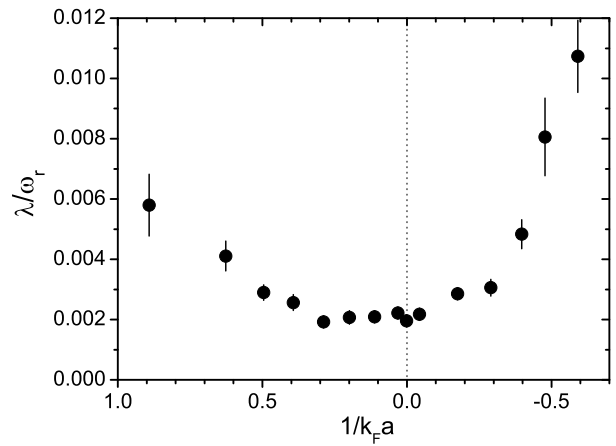}\\
  \caption{Lifetime of the angular momentum versus interaction
  parameter $1/k_Fa$ for $\epsilon=0.09$. The temperature for $1/k_Fa=0$ is $T/T_F=0.22(3)$.}\label{crossover}
\end{figure}

Figure \ref{crossover} shows the decay rate of the angular momentum
as a function of the interaction strength. The lifetime is largest
where the interaction is strongest and accordingly the relaxation
time is short. In addition to the two-body interaction strength,
pairing effects play an important role for the relaxation time
\cite{Riedl2008coo}. This might explain the higher decay rates for
$1/k_Fa<0$, where the pairing is weak, compared to the decay rates
for $1/k_Fa>0$, where the atoms are bound to molecules. Similar
behavior has been seen in \cite{Zwierlein2005vas} for the lifetime
of a vortex lattice. Note that Ref.~\cite{Zwierlein2005vas} also
reports a decrease of the lifetime in a narrow region around
$1/k_Fa=0$, which we do not observe for our trap parameters.

In summary the hydrodynamic behavior in the crossover region leads
to a very long lifetime of $L_z$.

\section{Conclusion}
In this work we have presented measurements on a strongly
interacting Fermi gas carrying angular momentum. The angular
momentum of the gas exhibits long lifetimes due to the deeply
hydrodynamic behavior of the normal state in such a system. We
investigated the decay rate of the angular momentum depending on the
ellipticity of the trapping potential for two different
temperatures. We find that the experimental results are in good
agreement with the expected behavior for a simple Boltzmann gas. The
dependence of the decay rate of the angular momentum on the
interaction strength in the BEC-BCS crossover region confirms that
collective motion is very stable as long as the interaction strength
is sufficiently large.

The long lifetime of the angular momentum in a rotating strongly
interacting Fermi gas allows us to further investigate rotational
properties both in the superfluid and normal phase in detail and
with high precision. Currently we investigate the moment of inertia
of the gas for different temperatures \cite{MOI}.


\begin{acknowledgments}
We acknowledge support by the Austrian Science Fund (FWF) within SFB
15 (project part 21) and SFB 40 (project part 4).
\end{acknowledgments}

\appendix
\section{}
To calculate the precession angle and the oscillation of the width
we assume that the frequency and damping rate for the $m=\pm2$
excitations are independent. For the damping of each excitation we
assume a exponential behavior. A superposition of the two
excitations results in the fit function for the precession angle
\cite{Bretin2003qoo}
\begin{eqnarray}\label{fitprecession}
&&\tan{(2(\phi-\phi_e))}= \nonumber\\
&&\frac{e^{-(\Gamma_+-\Gamma_-) t}\sin{(\omega_+ t + 2\phi_{0})} -
\sin{(\omega_- t + 2\phi_{0})}} {e^{-(\Gamma_+-\Gamma_-)
t}\cos{(\omega_+ t + 2\phi_{0})} + \cos{(\omega_- t + 2\phi_{0})}}
\end{eqnarray}
Here $\omega_\pm$ are the frequencies, $\Gamma_\pm$ are the damping
rates, $\phi_{0}$ is the initial angle for the two excitations and
$\phi_e$ is the precession angle resulting from the expansion of the
cloud. For the oscillation of the width difference $\Delta W$ we get
\begin{eqnarray}\label{fitwidth}
\Delta W^2  &=&  4 A e^{-(\Gamma_++\Gamma_-) t}
\cos^2{\left(\frac{(\omega_++\omega_-)}{2} t +
2\phi_{0}\right)} \nonumber\\
&+&A(e^{-\Gamma_+ t}-e^{-\Gamma_- t})^2,
\end{eqnarray}
where $A$ is the amplitude of the oscillation.


\section{}
Here we calculate the effect of the expansion of the cloud on the
precession angle. Assuming conservation of angular momentum during
the expansion, the rotation frequency $\Omega$ of the gas decreases
as the size of the cloud is increasing. We introduce an effective
precession time $t_e$ which accounts for the changing precession
angle $\phi$ during expansion. The total change of the precession
angle resulting from the expansion is given by
\begin{equation}\label{exp1}
    \phi_e=\int_0^{t_{\rm TOF}}\dot\phi(t) dt =\dot\phi(0) t_e,
\end{equation}
where $\dot\phi(0)$ is the precession frequency when the gas is
still trapped and $t_{\rm TOF}$ is the expansion time. Assuming that
also during the expansion $\dot\phi(t)=L_z/(2M r_{\rm rms}^2(t))$ is
still valid and inserting this into Eq.~\ref{exp1} we get
\begin{equation}\label{exp2}
    t_e=\int_0^{t_{\rm TOF}}{ r_{\rm rms}^2(0)/r_{\rm rms}^2(t)} dt.
\end{equation}
To calculate the relative increase of the cloud size during
expansion, $r_{\rm rms}^2(t)/r_{\rm rms}^2(0)$, we use the scaling
approach; see e.g.~\cite{Altmeyer2007doa}. For our experimental
parameters, $\omega_r=800$\,Hz and $t_{\rm TOF}=0.8$\,ms, we get an
effective precession time of $t_e=0.26$\,ms. This is shorter than
the typical precession time in the trap of $0.75$\,ms.


\begin{thebibliography}{42}
\bibitem[{rot()}]{rotatingframe}
\bibinfo{note}{$\epsilon'=(\omega_x'^2-\omega_y'^2)/(\omega_x'^2+\omega_y'^2)$,
  where $\omega_x'$ and $\omega_y'$ are the trap frequencies in the frame of
  the rotating potential.}
\bibitem[{rms()}]{rmswidth}
\bibinfo{note}{We determine $r_{\rm rms}$ at unitarity from the trap parameters
  using $E_F=2 M \omega_r^2 r_{\rm rms}^2 \sqrt{1+\beta}$ where we used the
  universal scaling parameter $\beta =-0.56$ \cite{Giorgini2008tou}. Note that
  this underestimates $r_{\rm rms}$ by a few percent because it does not take
  into account the finite temperature and the rotation of the gas. This does
  not affect the measurement of the lifetime of rotation as this depends on the
  relative change of $L_z$.}
\bibitem[{mod()}]{modefreqrotation}
\bibinfo{note}{Note that the frequency of quadrupole mode oscillation
  $\omega_q$ depends on the rotation frequency of the gas via
  $\omega_q^2=2\omega_r^2-\Omega^2$. This leads to a tiny shift of the maxima
  of $\Delta W^2$ but does not affect our measurement of $L_z$ within our
  experimental uncertainty.}
\bibitem[{tec()}]{technical}
\bibinfo{note}{This is the largest magnetic field where absorption images can
  be taken with our current experimental setup.}
\bibitem[{tem()}]{tempincrease}
\bibinfo{note}{To estimate the increase of the temperature through the decay of
  the rotation we assume that the rotation energy is completely converted into
  heat. In the experiments $L_z$ is well below $3\hbar$ which leads to a
  relative temperature increase of $\Delta T/T<0.02$ in the relevant
  temperature range. This is clearly below the uncertainty of our temperature
  measurement.}
\end{thebibliography}

\end{document}